\documentclass[12pt]{article}
\usepackage{latexsym}
\usepackage{amsfonts}

\newtheorem{thm}{Theorem}[section]
\newtheorem{cor}[thm]{Corollary}
\newtheorem{lem}[thm]{Lemma}
\newtheorem{pro}[thm]{Proposition}

\title{Probabilistic behavior of hash tables}
\author{Dawei Hong\footnote{D.~Hong and J.C.~Birget,
                       Dept.\ of Computer Science,
                       Rutgers University at Camden, Camden, NJ 08102,
                       USA, \{dhong, birget\}@camden.rutgers.edu}, \ \   
        Jean-Camille Birget\footnote{Supported in part by 
                              NSF grant DMS-9970471}, \ \ 
        Shushuang Man \footnote{ Dept.\ of Mathematics and Computer Science,
                   Marshall, MN 56258, USA, mans@southwest.msus.edu} 
       }
\date{}
\begin{document}
\maketitle

\begin{abstract}
We extend a result of Goldreich and Ron about estimating the collision
probability of a hash function. Their estimate has a polynomial tail.
We prove that when the load factor is greater than a certain constant,
the estimator has a gaussian tail. As an application we find an estimate of 
an upper bound for the average search time in hashing with chaining, 
for a particular user (we allow the overall key distribution to be different 
from the key distribution of a particular user). The estimator has a 
gaussian tail. 
\end{abstract}


\section{Introduction}

Hash tables have many applications in computer science \cite{CLRS}, \cite{Kn}.
We especially mention data bases, where hash tables are used for storing 
values of an attribute; see chapter 12 of \cite{SKS}.
Following the notation of \cite{CLRS}, a hash function is a function
$h: U \mapsto T$, where both the domain $U$ and the range $T$
are finite. Traditionally, $U$ is called the {\it key space} or the ``universe'', 
and elements $x \in U$ are called {\it keys}. The set $T$ is called the 
the {\it table}, and its elements are called the {\it table slots}. When
$h(x) = i$ we say that $h$ {\it hashes} the key $x$ into the slot $i$.
We shall denote by $n$ the cardinality of $T$ and we will simply assume that 
$T = \{1, \ldots, n\}$.
We assumed that $U$ is (very much) larger than $T$. 

\smallskip

We assume that a probability measure $q$ has been defined on $U$.
The probability of $S \ (\subset U)$ is denoted by ${\sf P}(S)$ 
$( \ = \sum_{x \in S} q(x))$. We also put the product measure on 
$U \times U$ and on $U^m$ (for any positive integer $m$); using 
the product measure amounts to saying that in a sequence of $m$ keys,
all the keys are {\it independent}.   

The probability on $U$ induces a probability measure on $T$:
The {\it probability that some key hashes to slot} $i \ (\in T)$ is \  
$p_i =  \sum_{x \in h^{-1}(i)} q(x)$ $= {\sf P}(h^{-1}(i))$. 

If two keys $x_1, x_2 \in U$ have the same hash value, these keys are said 
to {\it collide}. The {\it collision probability} of the hash function $h$ 
is defined to be \ ${\sf P}\{(x_1, x_2) \in U \times U : h(x_1) = h(x_2) \}$ 
(in short-hand this is denoted by ${\sf P}(h(x_1) = h(x_2))$). 
Here we use the product measure (i.e., keys are ``chosen independently'').
A {\it true collision} corresponds to keys $x_1, x_2 \in U$ such that 
$x_1 \neq x_2$ and $h(x_1) = h(x_2)$. 

Throughout this paper, $\|.\|$ denotes euclidean norm. It is straightforward 
to prove the following.
\begin{pro}
The collision probability of $h$ is equal to \  
$\sum_{i = 1}^n p_i^2 \ \ ( = \|p\|^2)$. 

\medskip

\noindent
Moreover, we always have \ $\sum_{i=1}^n p_i^2 \geq \frac{1}{n}$, 
and equality holds iff \  $p_i = \frac{1}{n}$ for all $i \in T$.
\end{pro}
Similarly, the probability that two independently chosen keys are equal is 
 \ $\sum_{x \in U} q(u)^2$. Hence, the probability of true collisions 
for $h$ is \ $\sum_{i = 1}^n p_i^2 \ - \ \sum_{x \in U} q(u)^2.$

Note that \ $\sum_{x \in U} q(u)^2$ \ will usually be very small
assuming that $U$ is very large (compared to $n$ and compared to the length
$m$ of key sequences used), and assuming that the probability distribution
$q$ on $U$ is not very concentrated. 
Therefore, the difference between the collision probability $\|p\|^2$ and
the probability of true collisions is usually quite small.

\medskip

In this paper we assume that collisions
are resolved by some form of {\it chaining}; i.e., all the keys that are 
hashed into one slot are stored in that slot.
For a hash table with chaining, we will simply assume that the search time
(for both successful or unsuccessful search) in a slot $i$ is proportional
to the number of keys stored in that slot; for simplicity, we simply identify 
search time in a slot and chain length in the slot.

\bigskip

\noindent {\bf Notation} ``$k_i(x)$'':  \     
Let $x = (x_1, \ldots, x_m)$ be a sequence of $m$ keys that are inserted into 
our hash table, and let $i$ be a slot ($i = 1, \ldots, n$).
We let $k_i(x)$ denote the number of keys (counted with multiplicities) 
inserted into slot $i$. (``With multiplicities'' means that if a key 
occurs several times in $x$ it is counted as many times as it occurs.)

Since in $k_i(x)$ we count keys with multiplicities, $k_i(x)$ is an upper bound
on the number of different keys stored in slot $i$.

\begin{pro}
For a sequence of keys $x = (x_1, \ldots, x_m)$ that are inserted, the 
number of collisions between keys in $x$ is  
$$\sum_{i = 1}^n \frac{k_i(x)(k_i(x) - 1)}{2}.$$
\end{pro}
The proof is straightforward. Recall that we count pairs of equal keys in 
the sequence $x$ as collisions. Since there are $\frac{m(m - 1)}{2}$ 
unordered pairs of key insertions in $x$, we call \    
$$\sum_{i = 1}^n \frac{k_i(x)(k_i(x) - 1)}{m(m - 1)}$$
the  {\em empirical collision probability} of $x$.
This concept, and its relation with the collision probability $\|p\|^2$,
were first studied by Goldreich and Ron \cite{GR}.  

\bigskip

In this paper we obtain two results, in the form of deviation bounds.  
(1) We give an estimation of the collision probability. 
(2) We give a deviation bound for an upper bound on the average search time.

In the second result we assume that the load factor is $> 9$
(see later for the exact assumptions).
Applications in data bases often lead to hash tables with
large load factor (\cite{SKS}, Chapter 12). 
We allow arbitrary key distributions.

\bigskip

\noindent {\bf Estimation of the collision probability}

\medskip

\noindent 
Our first result extends a result of Goldreich and Ron \cite{GR},
namely that \  $\sum_{i = 1}^n \frac{k_i(x)(k_i(x) - 1)}{m(m - 1)}$
 \ is a very good estimator for the collision probability $\|p\|^2$. 
How good the estimator is can be measured  by the relative error \    
$|\sum_{i=1}^n \frac{k_i(x)(k_i(x)-1)}{m(m-1)} \cdot \frac{1}{\|p\|^2}$
$ \ - \ 1|$. Their result, as well as ours, gives a deviation bound for 
this relative error. Goldreich and Ron \cite{GR} proved a polynominal deviation
bound for the estimator \ $\sum_{i=1}^n \frac{k_i(x)(k_i(x)-1)}{m(m-1)}$.
Their goal was to find sublinear-time algorithms for testing expansion
properties of bounded-degree graphs.

\begin{thm} {\rm (Goldreich and Ron \cite{GR}).} \
For all \ $\beta > 0$, $\lambda \geq 0$, if
$m = n^{1/2 + \beta + \lambda}$ then
$${\sf P} \left\{ \left| \sum_{i = 1}^{n}
\frac{k_i(x)(k_i(x) - 1)}{m(m - 1)}  \cdot \frac{1}{\|p\|^2} - 1 \right|
\leq \frac{3}{n^{\beta/2}} \right\}
 \ \geq \  1 - \frac{4}{9n^{\lambda}}.$$
\end{thm}
We extend the theorem of Goldreich and Ron as follows:

\smallskip

\begin{thm}
\label{OurResult1} \ 
For all \ $n > 24$, \ $\frac{1}{3} > \epsilon > 0$, \ $\delta > 0$, \ 
$s > 0$, if \ $m = \epsilon^{-2}n^{1 + \delta}$ \  we have 
$${\sf P} \left\{ \left| \sum_{i = 1}^n
\frac{k_i(x)(k_i(x) - 1)}{m(m - 1)} \cdot \frac{1}{\|p\|^2} - 1 \right| 
\leq \epsilon \left( 3 + \frac{6s}{n^{\delta/2}} +
\frac{5s^2\epsilon}{n^\delta} \right) \right\} 
 \ \geq \ 1 - \frac{10}{9} \, e^{-s^2 /4}.$$
\end{thm} 

\medskip

\noindent By taking $s = 2 \, n^{\delta/2}$, the expression \
$3 + \frac{6s}{n^{\delta/2}} + \frac{5s^2\epsilon}{n^\delta} $ \   
becomes \ $3 + 12 + 20 \, \epsilon$ $(< 22)$; here we use 
$\epsilon < \frac{1}{3}$. Therefore,
 
\begin{cor} \label{OurResult1_cor1} \  
For all \ $n > 24$, \ $\frac{1}{3} > \epsilon > 0$, \ $\delta > 0$, \
if \ $m = \epsilon^{-2}n^{1 + \delta}$ \  we have

\medskip

 \ \ \ \ \   \ \ \ \ \  
${\sf P} \left\{ \left| \sum_{i = 1}^n
\frac{k_i(x)(k_i(x) - 1)}{m(m - 1)} \cdot \frac{1}{\|p\|^2} - 1 \right|
\leq 22 \, \epsilon \right\}
 \ \geq \ 1 - \frac{10}{9} \, e^{-n^{\delta}}.$
\end{cor}

\smallskip

\noindent
Writing $\delta = \frac{\log C}{\log n}$, for $C > 1$, we obtain 
$n^{\delta} = C$, and $m = \epsilon^{-2}Cn$, i.e., the load factor is 
$L = C \, \epsilon^{-2}$. Therefore, 
\begin{cor} \label{OurResult1_cor2} \
For all \ $n > 24$, \ $\frac{1}{3} > \epsilon > 0$, \ and all $m$ such that 
$L = \frac{m}{n} > \epsilon^{-2} \ ( > 9)$ \  we have

$${\sf P} \left\{ \left| \sum_{i = 1}^n
\frac{k_i(x)(k_i(x) - 1)}{m(m - 1)} \cdot \frac{1}{\|p\|^2} - 1 \right|
\leq 22 \, \epsilon \right\}
 \ \geq \ 1 - \frac{10}{9} \, e^{- L \epsilon^2}.$$ 
\end{cor}

\smallskip

\noindent
Note that the assumptions of this Corollary impose the following relation
between $L$ and $\epsilon$: \ $\frac{1}{3} > \epsilon > \frac{1}{\sqrt{L}}$;
equivalently, $L = \frac{m}{n} > \epsilon^{-2} \ ( > 9)$.

\medskip

To compare with the result of Goldreich and Ron, let us pick 
$\epsilon = n^{- \beta / 2}$ in Corollary \ref{OurResult1_cor1}. Then 
$n^{1/2 + \beta + \lambda} = m = \epsilon^{-2}n^{1 + \delta}$ implies
$\delta = \lambda - \frac{1}{2}$. Hence our Corollary becomes:

\begin{cor} \label{OurResult1_cor3} \  
For all \ $n > 24$, \ $\beta > \frac{\log 3}{\log n}$, \  
$\lambda > \frac{1}{2}$, 
if \ $m = n^{1/2 + \beta + \lambda}$ \  we have
$${\sf P} \left\{ \left| \sum_{i = 1}^n
\frac{k_i(x)(k_i(x) - 1)}{m(m - 1)} \cdot \frac{1}{\|p\|^2} - 1 \right|
\leq \frac{22}{5} \, n^{- \beta / 2} \right\}
 \ \geq \ 1 - \frac{10}{9}e^{-n^{\lambda - \frac{1}{2}}}.$$
\end{cor}

\smallskip

Comparing \ref{OurResult1_cor3} with the theorem of Goldreich and Ron: 
Our theorem gives a much better deviation bound
(it is exponential, as opposed to the polynomial bound of Goldreich and Ron); 
but it applies only when the load factor $L$ is $> 9$ (whereas in the 
result of Goldreich and Ron, the load factor $L = n^{\beta + \lambda - 1/2}$ 
can be arbitrarily small, depending on $n$).

\bigskip

\noindent {\bf The average search time for a particular user}

\medskip

In order to analyze the efficiency of a hash table one considers the 
overall usage statistics of the keys (over all users). 
By ``user'' we mean a person or a process.
For every user we introduce a vector
$v = (v_1, \ldots, v_n)$, where $v_i$ is the frequency of the user's access 
(for search) to slot $i$. More precisely, $v_i$ is the number of searches at 
slot $i$, divided by the total number of searches in the table, for this user.
 Then $0 \leq v_i \leq 1$ and $\sum_{i=1}^n v_i = 1$.
We shall call $v$ the user's {\it access pattern}.
Traditional analysis of the average search time assumes that the accesses 
pattern of a user is the same as the key distribution (see e.g., \cite{CLRS}).

\smallskip

We let ${\rm AST}(v, x)$ denote the average search time for a user with access 
pattern $v$, under the condition that a sequence $x$ of $m$ independent keys
was previously inserted into the hash table. Clearly, we have the following 
upper bound:

\smallskip

  \ \ \ \ \ ${\rm AST}(v, x) \ \leq \ \sum_{i=1}^n v_i \cdot k_i(x)$.

\smallskip

\noindent The difference between ${\rm AST}(v, x)$ and 
$\sum_{i=1}^n v_i \cdot k_i(x)$ is caused by the possibility of 
pseudo-collisions. Here we are only concerned with upper bounds 
on ${\rm AST}(v, x)$, so we can use $\sum_{i=1}^n v_i \cdot k_i(x)$.

\smallskip

We write $m$ as $m = Ln$, where $L$ is called the {\it load factor}.
We do not assume that $L$ is a constant. Applying Theorem \ref{OurResult1} 
we show 
\begin{cor}
\label{OurResult2}
For all \ $n > 24$, \ $s > 0$, \ $L > 9$,  and $m = L n$ we have 
\[
{\sf P} \left\{ {\rm AST}(v, x) \leq  \ L \, n \|v\| \, \|p\| \, 
\sqrt{1+ \frac{3+6s}{\sqrt{L}} + \frac{5s^2}{L} } + 1 \right\}
 \ \geq \ 1 - \frac{10}{9}e^{-s^2 /4}. \]
\end{cor} 

\medskip

\noindent Noting that 
 \ $\sqrt{1+ \frac{3+6s}{\sqrt{L}} + \frac{5s^2}{L} }$ $<$ 
$1 + \frac{ 4s}{\sqrt{L}}$ \ and letting \ $\epsilon = \frac{s}{2\sqrt{L}}$ 
 \ we obtain 
\begin{cor}
\label{OurResult2_cor1} \ For all \ $n > 24$, \ $\epsilon > 0$, \ $L > 9$,  
and $m = L n$ we have           
\[
{\sf P} \left\{ {\rm AST}(v, x) \leq \ L \, n \|v\| \, \|p\| \,          
 (1 + 8\epsilon) + 1 \right\}
 \ \geq \ 1 - \frac{10}{9}e^{-L \epsilon^2}. \]
\end{cor} 

\noindent One notices that the probability bound is only interesting when 
$L$ is significantly larger than $\epsilon^{-2}$. Also, the error bound
is interesting only when $\epsilon$ is less than $1/8$; this means that 
the load factor has to be at least 100 for our results to be intersting.
In that sense, the results are theoretical, and show just what type of 
behavior to expect, up to big-O.  

In \cite{CLRS} (chapt.~12, exercise 12-3) the expected search time (for 
every user) was found to be $\Theta \left(L\right)$, under the assumption 
that both the key distribution and the distribution of user's accesses are 
uniform. 
Our Corollary implies that if $\|p\|^2 = \Theta \left(\frac{1}{n}\right)$ 
and $\|v\|^2 = \Theta \left(\frac{1}{n}\right)$
(which is much more relaxed than the assumption of a uniform distribution), 
then with exponentially high probability, the average search time is $O(L)$ 
for a user with access pattern  $v$.

\bigskip

\noindent {\bf Example 1} 

\smallskip

Suppose that a hash table, designed for a certain population of users, has 
collision probability \ $\|p\| \leq \frac{c}{\sqrt{n}}$ (for 
the overall population of users); $c$ is a positive constant.
The keys in the hash table are independent random samples.
Now consider an individual user who accesses a subset of cardinality
$\alpha \, n$ (where $0 < \alpha \leq 1$) of the $n$ slots of the hash table, 
with uniform probability $\frac{1}{\alpha n}$, and who does not access the 
other $(1 - \alpha) n$ slots of the hash table at all (i.e., those slots 
have probability 0 for this user).  Then the question is: What is the average 
search time for this user and this table, and what is the deviation bound?

Since the user accesses a fraction $\alpha$ of the slots uniformly, we have
$\|v\| = \frac{1}{\sqrt{\alpha n}}$. By Corollary \ref{OurResult2_cor1}, 

\smallskip

  \ \ \ \ \ ${\sf P}\{ {\rm AST}(v, x) \leq \ $
                   $\frac{cL}{\sqrt{\alpha}} \, (1 + 8\epsilon) + 1 \}$
 $ \ \geq \ 1 - \frac{10}{9}e^{-L \epsilon^2}$.

\smallskip

\noindent
So, the average search time is at most $1 + \frac{cL}{\sqrt{\alpha}}$ ,
with smaller error bound (namely \ $\frac{cL}{\sqrt{\alpha}} \, 8\epsilon$), 
and with probability close to 1 (namely \ 
 $1 - \frac{10}{9}e^{-L \epsilon^2}$). 

One observes that when the fraction $\alpha$ of the table used by the user
becomes smaller, the upper bound on the average search time for this user
increases, as does the error bound. This is not surprising; hashing works 
best when the keys are spread over the table as evenly as possible. 
Interestingly, our probability bound does not depend on $\alpha$. 

Some possible numerical values: For $c = 5$, \ $\alpha = 0.1$, 
 \ $\epsilon = 0.05$, $L = 1000$, we get \ 
${\rm AST}(v, x) \leq \ 15811 \pm 6324$, with probability at least
$1 - \frac{10}{9}e^{-L \epsilon^2}$ \ $= \ 0.909$.
For $c = 5$, \ $\alpha = 0.1$, \ $\epsilon = 0.05$, $L = 10000$, 
we get \ ${\rm AST}(v, x) \leq \ (1.58 \pm 0.64) \cdot 10^5$, with 
probability at least $1 - 1.54 \cdot 10^{-11}$.

\bigskip

\noindent {\bf Example 2}

\smallskip

Let us consider the situation in which a query consists of two subqueries,
$Q_1$ and $Q_2$. This happens very commonly (e.g., in a ``three-tier
architecture''); see \cite{SKS}.
The two subqueries  can be viewed as two users with 
access patterns $v^{(1)}$ and $v^{(2)}$. Assume, for this example, that 
each of $Q_1$ and $Q_2$ behaves like the user in Example 1 above.
In particular, for $Q_i$ $(i = 1, 2)$ we have \ 
$\|v^{(i)}\| = \frac{1}{\sqrt{\alpha_i n}}$,  and 

\smallskip

  \ \ \ ${\sf P}\{ {\rm AST}_i(v^{(i)}, x) \leq \ $
                   $\frac{cL}{\sqrt{\alpha_i}} \, (1 + 8\epsilon) + 1 \}$
 $ \ \geq \ 1 - \frac{10}{9}e^{-L \epsilon^2}$.

\smallskip

\noindent Hence, for the combined query the average search time is a 
weighted sum \ 

\smallskip

  \ \ \ ${\rm AST} = w_1 \cdot {\rm AST}_1 + w_2 \cdot {\rm AST}_2$, 
  \ \ \ with $w_1 + w_2 = 1$. 

\smallskip

\noindent Let \ $a_i = \frac{cL}{\sqrt{\alpha_i}} \, (1 + 8\epsilon) + 1$. Then \   

\smallskip
   
${\sf P}\{ {\rm AST} \leq w_1a_1 + w_2a_2 \} \ \geq \ $ 
${\sf P}\{ {\rm AST}_1 \leq {\rm max}\{a_1,a_2\}, \ $
          ${\rm AST}_2 \leq {\rm max}\{a_1,a_2\}\} $

\smallskip

$ \geq \ 1 - 2 \, \frac{10}{9}e^{-L \epsilon^2}$.

\smallskip

\noindent Therefore, the average search time AST$(v^{(1)}, v^{(2)}, x)$
 of the combined query satisfies 

\smallskip

 \ \ \ ${\sf P}\{ {\rm AST}(v^{(1)}, v^{(2)}, x) \leq \ $
 $\frac{cL}{\sqrt{{\rm min}\{\alpha_1,\alpha_2\}}} \, (1+ 8\epsilon) +1 \}$
 $ \ \geq \ 1 - \frac{20}{9}e^{-L \epsilon^2}$.

\bigskip

\noindent Hence, when the load factor is large (compared to $\epsilon^2$) we 
obtain a very reliable upper bound on the average search time for the 
combined query. The knowledge of this upper bound enables various processes 
(that wait for the completion of this query) to be scheduled in a predictable 
way. 

The constants in our results are rather large. This is due to the generality
of our results. In a precise practical situation, our results could be used
for the format of the probabilistic behavior, with constants to be 
determined empirically.

\bigskip

The next section contains the proofs of our theorems.

\section{Proofs}

\subsection{A deviation bound for the empirical
collision probability: Proof of Theorem \ref{OurResult1} }
 
Our main technique will be  Talagrand's isoperimetric theory, developed 
by Talagrand in the mid 1990s \cite{Ta}. It has had a profound
impact on the probabilistic theory of combinatorial optimization \cite{St}
(see Sections 6 - 13 of \cite{Ta} and chapter 6 of \cite{St}).
 
Let $(\Omega, \mu)$ be a probability space, and let $(\Omega^m, \mu^m)$
be the product space. For $x \in \Omega^m$ and $A \subset \Omega^m$, 
Talagrand's convex distance $d_T(x, A)$ is defined by
\[     d_T(x, A) = \sup_{\alpha} \left \{z_\alpha  = \inf_{y \in A}
\left\{\sum_{j = 1}^m \alpha_i \ {\bf 1}(x_j \not= y_j) \right\} \ : \ 
 \alpha = (\alpha_1, \ldots, \alpha_m), \ \sum_{j = 1}^m \alpha_j^2 \leq 1  
\right\},
\]
where $x$ = $(x_1, \ldots , x_m)$,  $y$ = $(y_1, \ldots , y_m)$. 
Here, ${\bf 1}(x_i \not= y_i)$ = 1 if $x_i$ $\not=$ $y_i$, and it is 0
otherwise.
 
\begin{thm}
\label{Ta}
{\rm (Talagrand 1995)} \ 
For every $A \subset \Omega^m$ with $\mu^m(A) > 0$, we have
$$\int_{\Omega^m} \exp \left(\frac{1}{4} d_T (x, A)^2 \right) d \mu^m(x)
  \  \leq \   \frac{1}{\mu^m(A)},$$
and consequently, we have for all $s > 0$, 
$${\sf P} \left \{ d_T(x, A) \geq s \right \} \ \leq \  
\frac{e^{-s^2/4}}{\mu^m(A)}.$$
\end{thm}

\medskip

\noindent 
To apply Talagrand's theorem to our situation we define a set 
$A \subseteq U^m$ by 
\[      A \ = \  \left\{y \in U^m  \ : \  \left|
\sum_{i=1}^n \frac{k_i(y)(k_i(y) - 1)}{m(m - 1)} \cdot
\frac{1}{\|p\|^2} - 1 \right|
\leq 3\epsilon
\right\}.        \]

\begin{lem}
\label{SubsetA} \ 
For all $n > 24$ we have \  ${\sf P}(A) \geq \frac{9}{10}.$
\end{lem}
{\bf Proof.} \ Recall that $m = \epsilon^{-2}n^{1 + \delta}$  with
$\frac{1}{3} > \epsilon > 0$, \ $\delta > 0$. 
Letting $\beta = \frac{- 2 \log \epsilon}{\log n}$ and
$\lambda = 1/2 + \delta$, we rewrite $m$ as $n^{1/2 + \beta + \lambda}$.
Then the lemma follows from Theorem of Goldreich and Ron.
 \ \ \ $\Box$
 
\bigskip
 
\noindent For every $s > 0$ we define a set $C_s \subseteq U^m$ by 
$$C_s = \{ x \in U^m : d_T(x, A) < s \}.$$
By Theorem \ref{Ta} and Lemma \ref{SubsetA} we have for all $n > 24$
and all $s > 0$
\begin{equation}
\label{SubsetC}
{\sf P}(C_s) \geq 1 - \frac{10}{9}e^{-s^2/4}.
\end{equation}
\begin{lem}
\label{ExpandingA} \ 
For every $x = (x_1, \ldots, x_m) \in C_s$ there is 
$y = (y_1, \ldots, y_m) \in A$ such that 
 
  $$\sum_{j = 1}^m {\bf 1}(x_j \not= y_j) \ \leq \ sm^{1/2}.$$
\end{lem}
{\bf Proof.} \ Assume, by contradiction, that there is $x \in C_s$ such that 
for all $y \in A$, \ \ 

\smallskip

$\sum_{j = 1}^m {\bf 1}(x_j \not= y_j) > sm^{1/2}.$

\smallskip

\noindent Now, if we take $\alpha = (\alpha_1, \ldots , \alpha_m)$
$ = (m^{-1/2}, \ldots , m^{-1/2})$ in the definition of the Talagrand 
distance $d_T$, 
the inequality above implies  \ $d_T(x, A_1) \geq s$. But since $x \in C_s$, 
we also have $d_T(x, A_1) < s$, a contradiction.  \ \ \ $\Box$
 
\bigskip
 
Recall that for any $x = (x_1, ..., x_m)$, $y = (y_1, ..., y_m)$ $\in U^m$,
we defined $k_i(x)$ (resp. $k_i(y)$) to be the number of the keys (with
multiplicity) that are hashed into the slot $i$ for input sequence $x$, resp.
$y$. We define integers $s_i$ ($1 \leq i \leq n$) by
$$k_i(x) = k_i(y) + s_i.$$
\begin{lem}
\label{ExpandingA0} \ 
For all $x, y \in U^m$, \ 
$$\sum_{i = 1}^n |s_i| \ \leq \ 2 \sum_{j = 1}^m {\bf 1}(x_j \not= y_j).$$
\end{lem}
{\bf Proof.} \ We prove the lemma by induction on \ 
$\sum_{i = 1}^m {\bf 1}(x_i \not= y_i)$.
 
\smallskip
 
\noindent {\bf (0)} \ \ $\sum_{j = 1}^m{\bf 1}(x_j \not= y_j) = 0$: 

\smallskip

\noindent Then we have $x_j = y_j$ for all $j = 1, \ldots, m$, and hence,
$k_i(x) = k_i(y)$ for all $i = 1, \ldots, n$. Thus, we have
$\sum_{i = 1}^n |s_i| = 0$, finishing the base case.

\smallskip
 
\noindent {\bf (Inductive step)} \ \ Assume \  
$\sum_{j = 1}^m{\bf 1}(x_j \not= y_j) > 0$:

\smallskip

\noindent Without loss of generality we assume that $x_m \not= y_m$. 
Now, consider $\bar x = (x_1, \ldots , x_{m - 1}, y_m)$. We write
$k_i(\bar x) = k_i(y) + \bar {s_i}$ for $i = 1, \ldots, n$.
By the induction hypothesis we have
\begin{equation}
\label{ExpandingA0_1}
\sum_{i=1}^n |\bar s_i| \ \leq \ 2 \sum_{j=1}^m {\bf 1}(\bar x_j \not= y_i).
\end{equation}
Since $x$ differs from $\bar x$ only in its last component, we either have 
$h(x_m) = h(y_m)$, in which case 
$\bar s_i = s_i$ for all $i = 1, \ldots , n$.  Or we have 
$h(x_m) \neq h(y_m)$; let $i_1 = h(x_m)$ and $i_2 = h(y_m)$. Then 
 \ $\bar s_{i_1} = s_{i_1} + 1$, \  $\bar s_{i_2} = s_{i_2} - 1$, and
$\bar s_i = s_i$ for all $i \in \{1, \ldots , n \} \setminus \{i_1, i_2\}$.
In both cases, 
\begin{equation}
\label{ExpandingA0_2}
\left| \sum_{i = 1}^n |\bar {s_i}| -
\sum_{i = 1}^n |s_i| \right| \leq 2.
\end{equation}
On the other hand, 
$$\sum_{j = 1}^m {\bf 1}(x_j \not= y_j)  \ = \  
\sum_{j = 1}^m {\bf 1}(\bar x_j \not= y_j) + 1.$$
Combining this, (\ref{ExpandingA0_1}), and (\ref{ExpandingA0_2}), completes 
the proof for the inductive step.  \ \ \ $\Box$
 
\begin{lem}
\label{ExpandingA1} \  
For every $x \in C_s$ there is $y \in A$ such that for all $n > 24$, 
$0 < \epsilon < 1/3$, $s > 0$, and $m = \epsilon^{-2}n^{1+\delta}$, 
 we have 
$$\left|
\sum_{i = 1}^{n} \frac{k_i(x)(k_i(x) - 1)}{m(m - 1)} -
\sum_{i = 1}^{n} \frac{k_i(y)(k_i(y) - 1)}{m(m - 1)}
\right| \ \leq \ \epsilon \|p\|^2 \left(
\frac{6s}{n^{\delta/2}} + \frac{5s^2\epsilon}{n^{\delta}} \right).$$
\end{lem}
{\bf Proof.} \ For any fixed $x \in C_s$ we take $y \in A$ according to 
Lemma \ref{ExpandingA}.  That is,
\begin{equation}
\label{ExpandingA1_1}
\sum_{j = 1}^m {\bf 1}(x_j \not= y_j) \ \leq \ s m^{1/2}.
\end{equation}
As in the proof for Lemma \ref{ExpandingA0} we use the notation 
$k_i(x)$, $k_i(y)$, and $s_i$ \ ($i = 1, \ldots , n$).
We will leave the common denominator $m(m-1)$ out of the computations 
until the end:

\medskip

$| \sum_{i=1}^n k_i(x)(k_i(x) - 1) \ - \ \sum_{i=1}^n k_i(y)(k_i(y)-1) |$

\medskip

$= |\sum_{i=1}^n (k_i(y) +s_i)(k_i(y)+s_i-1) \ -  \ $
$\sum_{i=1}^n k_i(y)(k_i(y)-1) |$

\medskip

$= |\sum_{1\leq i \leq n, \, k_i(y) \geq 1} \ $
  $[(k_i(y) +s_i)(k_i(y)+s_i-1) - k_i(y)(k_i(y)-1)] $

\smallskip

 \hspace{3in}  $ \ + \ \sum_{1\leq i \leq n, \, k_i(y) = 0} \ s_i(s_i -1) |$ 

\medskip

$\leq \ \sum_{1\leq i \leq n, \, k_i(y) \geq 1} \ 2\, |s_i|(k_i(y)-1)$
  $ \ + \ \sum_{1\leq i \leq n, \, k_i(y) \geq 1} \ (s_i^2 + |s_i|)$ 
  $ \ + \ | \sum_{1\leq i \leq n, \, k_i(y) =0} \ s_i(s_i -1)|$

\medskip

$\leq \ \sum_{1\leq i \leq n, \, k_i(y) \geq 1} \ 2\, |s_i|(k_i(y)-1)$ 
$ \ + \ \sum_{i=1}^n (s_i^2 + |s_i|).$ 

\medskip 

\noindent By the Cauchy-Schwarz inequality, this is bounded by 

\smallskip

$\leq \ 2 \ (\sum_{1\leq i \leq n, \, k_i(y) \geq 1} \ s_i^2)^{1/2}$
       $(\sum_{1\leq i \leq n, \, k_i(y) \geq 1} \ (k_i(y)-1)^2)^{1/2}$ 
   $ \ + \ \sum_{i=1}^n (s_i^2 + |s_i|)$ 

\medskip

$\leq \ 2 \ (\sum_{i=1}^n s_i^2)^{1/2}$ 
  $(\sum_{i=1}^n k_i(y)(k_i(y)-1)^2)^{1/2}$ 
 $ \ + \ \sum_{i=1}^n (s_i^2 + |s_i|).$ 

\medskip 

\noindent By Lemma \ref{ExpandingA0} and (\ref{ExpandingA1_1}) we have
\begin{equation}
\label{ExpandingA1_4}
\sum_{i=1}^n s_i^2 \ \leq \ \left(\sum_{i=1}^n |s_i| \right)^2 \ \leq \  
\left(2 \sum_{j=1}^m {\bf 1}(x_j \not= y_j) \right)^2 \ \leq \ 4 s^2 m.
\end{equation}
Since $y \in A$ we have
$$\sum_{i=1}^n \frac{k_i(y)(k_i(y) - 1)}{m (m - 1)} \leq
\|p\|^2 \left(1 + 3\epsilon \right).$$
Hence, by all the above:
$$\left| \sum_{i=1}^n \frac{k_i(x)(k_i(x) - 1)}{m(m - 1)} -
\sum_{i=1}^n \frac{k_i(y)(k_i(y) - 1)}{m(m - 1)} \right|$$
$$\leq \ \frac{4s}{(m - 1)^{1/2}} \cdot \|p\|
\left(1+ 3\epsilon \right)^{1/2} + \frac{4s^2}{m-1} + \frac{2s}{m^{1/2}(m-1)}.$$

\noindent By calculating, and using the fact that $\|p\|^2 \geq \frac{1}{n}$, 
$0 < \epsilon < 1/3$, and $m = \epsilon^{-2}n^{1 + \delta}$, we find the 
following upper bound for \ 
$\left|
\sum_{i = 1}^{n} \frac{k_i(x)(k_i(x) - 1)}{m(m - 1)} -
\sum_{i = 1}^{n} \frac{k_i(y)(k_i(y) - 1)}{m(m - 1)}
\right| \ $: 
\[
  \frac{s\epsilon}{n^{\delta/2}} \ \|p\|^2 \ 
 \frac{4(1+3\epsilon)^{1/2}n^{1/2}}{(n - \epsilon^2 n^{-\delta})^{1/2} } 
 \ + \ 
\frac{s\epsilon}{n^{\delta/2}} \ \|p\|^2 \  
  \frac{2 \epsilon^2n}{n^{1/2} (n^{1+\delta} - \epsilon^2)}
\ + \  
 \frac{s^2\epsilon^2}{n^{\delta/2}} \ \|p\|^2 \  
 \frac{4n}{n^{1+\delta} - \epsilon^2}
\]
\noindent Combining this and using $n > 24$ we obtain the upper bound \ 
$$\epsilon \|p\|^2 (\frac{6s}{n^{\delta/2}} + 
\frac{5s^2 \epsilon}{n^{\delta}}).$$
$\Box$

\bigskip
 
\noindent
{\bf Proof of Theorem \ref{OurResult1}.} \  
The theorem follows from the definition of $A$, inequality (\ref{SubsetC}), 
and Lemma \ref{ExpandingA1}. \ \ \ $\Box$

 
\subsection{Average search time for a particular user}

\noindent
{\bf Proof of Corollary \ref{OurResult2}.} \   
Recall that the average search time AST$(v,x)$ is bounded from above by \
$\sum_{i = 1}^n v_i \cdot k_i(x)$.
In Theorem \ref{OurResult1} let us write $m = L_1 L_2 \, n$, and choose  
$$\epsilon = \frac{1}{\sqrt L_1}~~{\rm and}~~\delta = \frac{\log L_2}{\log n}.$$
Note that for all $i$, \ 

\smallskip

$k_i(x) - 1 \leq \sqrt{k_i(x)(k_i(x) - 1)}$ \   

\smallskip

\noindent since the left side is 0 when $k_i(x) = 0$ or 1. Therefore,  

\smallskip

${\rm AST}(x,v) \leq \ $
$\sum_{i=1}^n v_i \cdot k_i(x) \ = \ \sum_{i=1}^n v_i (k_i(x) - 1) + 1 \  $
$\leq \ \sqrt{\sum_{i=1}^n v_i^2} \sqrt{\sum_{i=1}^n (k_i(x) - 1)^2} +1$

\medskip

$\leq \ \sqrt{\sum_{i=1}^n v_i^2} \sqrt{\sum_{i=1}^n k_i(x)(k_i(x) - 1)} + 1$
$ \ \leq \ \|v\| \, \|p\| \, m(m-1)$
  $\sqrt{\sum_{i=1}^n \frac{k_i(x)(k_i(x)-1)}{m(m-1)} \, \frac{1}{\|p\|^2}}$.

\medskip

\noindent The corollary follows from this and Theorem \ref{OurResult1}.
 \ \ \   $\Box$   

\bigskip

\noindent
{\bf Remark.} \ Our proof method depends crucially on Talagrand's theorem.
Many readers, more familiar with techniques like the Chernoff bound, or more
generally, the Hoeffding inequality for martingale differences (from which 
the Chernoff bound follows directly), may wonder whether these simpler 
techniques don't work here. In order to apply Hoeffding's inequality we 
could view $\sum_{i=1}^n v_i \cdot k_i(x)$ as a weighted sum of the random
variables $k_i(x)$; to apply Hoeffding one needs to bound $|k_i(x)|$, but 
we don't have good bounds a priori; finding good bounds on $|k_i(x)|$ seems
harder and less promising than our method, based on Talagrand's theorem.
See, e.g., Michael Steele's book \cite{St}, which discusses the advantages 
of applying Talagrand's theorem at length.


\end{document}